\documentclass{aastex63}

\received{July 25, 2019}
\revised{October 10, 2019}
\accepted{October 18, 2019}
\submitjournal{ApJ}

\shorttitle{Multiple populations in globular clusters}
\shortauthors{Jang, Kim, \& Lee}

\begin{document}

\title{Multiple populations in globular clusters:\\Unified efforts from stellar evolution and chemical evolution models}

\correspondingauthor{Sohee Jang, Jenny J. Kim, Young-Wook Lee}
\email{sohee.jang@yonsei.ac.kr, kim@uni-heidelberg.de, ywlee2@yonsei.ac.kr}



\author[0000-0002-1562-7557]{Sohee Jang}
\affiliation{Center for Galaxy Evolution Research and Department of Astronomy,\\Yonsei University, Seoul 03722, Korea}


\author[0000-0002-0432-6847]{Jenny J. Kim}
\affiliation{Center for Galaxy Evolution Research and Department of Astronomy,\\Yonsei University, Seoul 03722, Korea}
\affiliation{Astronomisches Rechen-Institut, Zentrum für Astronomie der Universität Heidelberg,\\
Mönchhofstraße 12-14, 69120 Heidelberg, Germany}

\author[0000-0002-2210-1238]{Young-Wook Lee}
\affiliation{Center for Galaxy Evolution Research and Department of Astronomy,\\Yonsei University, Seoul 03722, Korea}







\begin{abstract}

Recent stellar evolution models for globular clusters (GCs) in multiple population paradigm suggest that horizontal-branch (HB) morphology and mean period of type ab RR Lyrae variables are mostly determined by  He \& CNO abundances and relative ages for subpopulations. These parameters are also provided by chemical evolution models constructed to reproduce the Na-O anti-correlation. Therefore, a consistency check is possible between the synthetic HB and chemical evolution models. Furthermore, by combining them, a better constraint might be attained for star formation history and chemical abundances of subpopulations in GCs. We find, from such efforts made for four GCs, M4, M5, M15, and M80, that consistent results can be obtained from these two independent studies. In our unified model, He and Na abundances gradually increase over the generation, and therefore, the various extensions observed in both HB morphology and Na-O chemical pattern depend on the presence of later generation stars after the second generation. It is schematically shown that this observed diversity, however, would not be naturally explained by the models requiring dilution. Further spectroscopic observations are required, for metal-poor GCs in particular, to obtain a more detailed constraint from this approach.\\ 

\end{abstract}

\keywords{stars: horizontal-branch --- stars: abundances --- globular clusters: general --- Galaxy: halo} 


\section{Introduction} \label{sec:intro}

It is now well established from the observations performed during the past two decades that most globular clusters (GCs) host multiple stellar populations with different He and light element abundances \citep[][and references therein]{Gra12,Ren15,Bas18}. Photometric observations have convincingly shown clear signatures of multiple populations in all evolutionary stages on the color-magnitude diagram (CMD) \citep[][and references therein]{Lee99,Bed04,Han09,Pio15,Mil17}. Stellar evolution models have successfully reproduced these discrete features by employing multiple generations/subpopulations with different He and N abundances. Since horizontal-branch (HB) morphology is most sensitive to the variation in He abundance, a number of these studies were carried out to reproduce the observed properties of HB in the CMD \citep{Dan04,Lee05,Gra10,Joo13,Kun13}. In particular, \citet{Jan14} and \citet{Jan15} have further shown that, in order to reproduce the mean period of type ab RR Lyrae stars ($\left<P_{\rm ab}\right>$), together with the HB morphology, most of GCs are required to host mildly He-enhanced second generation stars (G2) and more He-rich third and later generations (G3, G4,...). In some cases, such as M15, an enhanced CNO abundance for G2 and a relatively large age difference between the first generation stars (G1) and G2 ($\Delta$t(G1 - G2)) are also suggested. These studies employing the synthetic HB models can provide a minimum number and He abundances of subpopulations required to reproduce the HB morphology from red to blue ends. In order to understand the more detailed chemical composition and star formation history for subpopulations in a GC, however, additional information from spectroscopy is required.\\

Spectroscopic investigations have also revealed clear evidence of multiple stellar populations through a star-to-star variation in light elements, most notably the Na-O anti-correlation \citep[][and references therein]{Car09b,Gra15,Bra17}. Recent theoretical studies of chemical evolution in GCs have been carried out mostly to reproduce this unique Na-O pattern. Among several scenarios suggested, the most widely accepted view is that the chemical evolution in GCs is dictated by the ejecta from asymptotic giant branch (AGB) stars and/or the wind of massive stars (WMS) with a negligible contribution from supernovae (SNe) \citep{Dan04,Dec07a,Dec07b,Der10,Kra13,Dan16}. Most of the previous models, however, assume that SN feedback would basically wipe out the ambient gas in a proto-GC, and therefore the G2 had to form either before the SN explosion from the gas polluted by WMS or after the SN from the gas enriched by AGB. In our recent chemical evolution model \citep{Kim18}, we have released this constraint and assume that SN blast waves undergo blowout without expelling the gas in a proto-GC. In this new scenario, multiple populations in GCs are best explained by multiple episodes of star formation and enrichment. The observed Na-O anti-correlations in most GCs can be reproduced and the “mass budget problem” is mostly resolved in this new model without ad hoc assumptions on star formation efficiency (SFE), initial mass function (IMF), and preferential loss of G1 stars. Because star formation timescale ($\Delta$t) between the subsequent generations is the main factor determining the chemical evolution of a GC in our model, once a star formation history reproducing the observed Na-O anti-correlation is settled, He and CNO abundances corresponding to Na and O abundances of each subpopulation are simultaneously obtained. However, measuring the He and CNO abundances from spectroscopy is still a rather challenging task. Fortunately, they can be independently determined from synthetic HB model and, therefore, can be used to verify the results from chemical evolution model. There has been no attempt in multiple population studies, however, to directly combine the outcomes from stellar evolution models for HB and chemical evolution models. The purpose of this paper is to see whether a consistent result can be obtained from these two independent models, and, by combining them, to place a better constraint on star formation history and chemical evolution in a GC with multiple populations.

\section{Stellar evolution and chemical evolution models for four globular clusters} \label{sec:model}

Our synthetic CMDs are based on the most updated Yonsei-Yale ($\rm Y^2$) HB evolutionary tracks and isochrones with enhanced He and N abundances \citep{Lee15}. Following our recent investigations for the HB morphology of GCs with multiple populations, the Reimers mass-loss parameter $\eta$ required in the construction of synthetic HB models was employed to be 0.50, and the mass dispersion on the HB was adopted to be 0.010 $\rm M_{\sun}$ for each subpopulation \citep{Jan14,Jan15,Lee16}. The fundamental periods of RR Lyrae variables are also calculated in the synthetic HB modeling from the period-mean density relation \citep[see][]{Lee90}. Chemical evolution models presented in this paper are constructed based on the new prescriptions described in \citet{Kim18}. As described above, the most important assumption of this model is that SN ejecta escapes without expelling the pre-enriched ambient gas in a proto-GC. This is predicted by recent hydrodynamic simulations with more realistic treatment on the proto-GC environment and is also supported by some infrared and radio observations \citep{Ten15,Oey17,Tur17,Sil17,Sil18,Coh18}. Based on our previous stellar evolution modeling \citep{Lee05,Jan14,Jan15}, star formation and enrichment beyond G2 are allowed to continue to G3 and later generations, and thus gas becomes more and more enriched by processed materials. The readers are referred to our previous investigations for the details of the techniques and prescriptions in the construction of synthetic HB model and chemical evolution model \citep{Jan15,Kim18}. \\

In our modeling for four GCs, M4 (NGC 6121), M5 (NGC 5904), M15 (NGC 7078), and M80 (NGC 6093), the metallicity is adopted by fitting the isochrones with the observed red giant branch locus in CMD within the metallicity scale uncertainty of $\pm$ 0.2 dex from the observed [Fe/H] value listed in \citet{Har96} or \citet{Lee94}. For synthetic HB model, the He abundance has the greatest effect in reproducing the observed HB morphology and $\left<P_{\rm ab}\right>$. As described in \citet{Joo13} and \citet{Jan15}, He contents (Y’s) of subpopulations, together with CNO abundances, can be obtained by reproducing the observed CMD and $\left<P_{\rm ab}\right>$ to within the fitting error ($\sigma_{\rm Y}$ $\approx$ 0.01 and $\sigma_{\rm [CNO/Fe]}$ $\approx$ 0.2 dex). In our chemical evolution modeling, star formation timescale ($\Delta$t’s between G1 \& G2, G2 \& G3,…) is the only variable in reproducing the observed Na-O anti-correlation once parameters for SFE and IMF slope (s) are fixed within the ranges suggested by observations and theoretical studies (SFE = 0.4 - 0.6 \& s = 1.8 - 2.0; see \citealt{Kim18}). Therefore, in principle, chemical evolution models for a given GC can provide $\Delta$t’s between subpopulations to within the fitting error of $\pm$ 0.04 Gyr (see Table~\ref{tab1}).  In combining these two models, synthetic HB models for a given GC first provide initial estimates of He and CNO abundances and the minimum number of subpopulations to chemical evolution modeling. Then, within the uncertainties of He and CNO abundances obtained from synthetic HB models, star formation time scales and the number of subpopulations required to reproduce the Na-O anti-correlation are estimated in chemical evolution model. Note that, since He, Na, and O are supplied by similar sources in a GC (see Table 2 of \citealt{Kim18}), in chemical evolution modeling, a combination of Na and O abundances is obtained that corresponds to the specific value of He abundance. The synthetic HB model is, then, constructed again with these $\Delta$t’s and the number of subpopulations suggested by the chemical evolution model. These processes are repeated until the two models agree to within $\Delta$Y $\approx$ 0.01, $\Delta$t $\approx$ 0.04 Gyr, and $\Delta$ [CNO/Fe] $\approx$ 0.03 dex. As a result, we can simultaneously reproduce the Na-O anti-correlation, HB morphology, and $\left<P_{\rm ab}\right>$ with relatively low uncertainties ($\sigma_{\rm [Na/Fe]}$ $\approx$ 0.06, $\sigma_{\rm [O/Fe]}$ $\approx$ 0.08, $\sigma_{\rm HB type}$ $\approx$ 0.03, $\sigma_{\rm <P_{ab}>}$ $\approx$ 0.015 day). Table~\ref{tab1} shows the best-fit parameters from this unified endeavor from the synthetic HB models and the chemical evolution models for the four GCs considered in this Letter. It is encouraging to see that our predictions for the enhancement in He between G1 and G2+, $\Delta$Y = 0.02 (M4) - 0.07 (M15), are roughly comparable to the maximum internal He variations estimated from the chromosome map \citep[$\Delta$Y = 0.014 - 0.07; see Table 4 of][]{Mil18}. Our models also predict some enhancement in [CNO/Fe] for the two Oosterhoff group II GCs ($\Delta$[CNO/Fe] = 0.4 - 0.6 for M80 and M15). While spectroscopic observations for the CNO sum are still required for these GCs, similar variations in CNO abundance  have been reported in some GCs \citep{Alv12,Mar12,Yon15}.\\

\begin{figure}
\centering
\epsscale{1.05}
\plotone{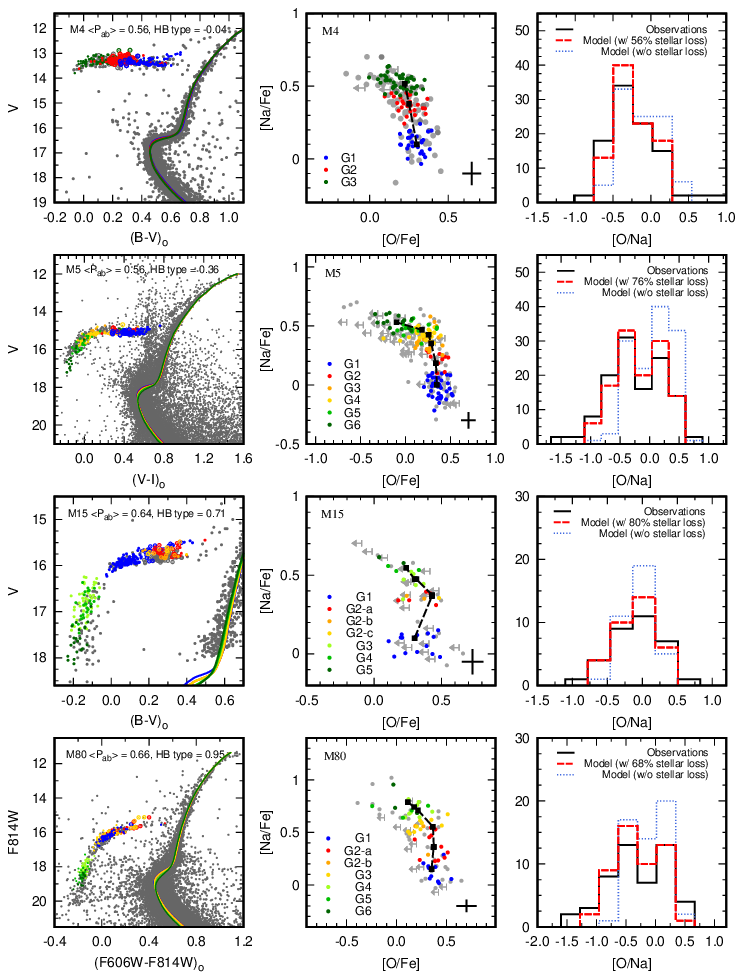}
\caption{Comparison of our synthetic HB and chemical evolution models with photometric and spectroscopic observations for M4, M5, M15, and M80 \citep[data in gray color from][]{Moc02,Ste14,Via13,Buo85,Bin84,And08,Car09a,Car09b,Car15}. The left panels compare the synthetic HB models and isochrones (colored circles and lines) with the observed CMDs. The values for $\left<P_{\rm ab}\right>$ and HB type are from our models. The middle and right panels are our chemical evolution models compared with the observed Na-O anti-correlations and [O/Na] histograms (see the text for the details). In the middle panels, our model predictions with/without observational errors are depicted by colored circles and filled black squares. \label{fig1}}
\end{figure}

Figure~\ref{fig1} compares our best-fit models for the four GCs with the observations. The left panels compare our synthetic HB models and isochrones with the observed CMDs. The data adopted for the photometric observations are from \citet{Moc02} and \citet{Ste14} for M4, \citet{Via13} for M5, \citet{Buo85} and \citet{Bin84} for M15, and \citet{And08} for M80. The middle and right panels show our chemical evolution models compared with the observed Na-O anti-correlations and [O/Na] histograms\footnote{In our chemical evolution models, for the best-fit with the observations, some preferential loss of earlier generations are required (see the right panels of Figure~\ref{fig1} and Table~\ref{tab1}), which is supported by a recent simulation on the dynamical disruption of proto-GCs \citep{Rei18}. Note that this preferential loss is an order of magnitude smaller than those required in other models.}. The spectroscopic data used for this comparison are from  \citet{Car09a,Car09b,Car15}. Two upper panels of Figure~\ref{fig1} show our models for the relatively metal-rich Oosterhoff group I GCs (M4 and M5). In our models for these GCs, subpopulations belonging to later generations are gradually placed on the bluer parts of the HB over the generation. Our synthetic HB models show that type ab RR Lyraes in these GCs are produced mainly by He-normal G1, producing relatively short $\left<P_{\rm ab}\right>$ (see Table~\ref{tab1}). It is evident that M5 is far more extended than M4 both on the HB and the Na-O chemical pattern. Our models suggest that the star formation episode of M5 is similar to that for M4 until the formation of G3, but additional generations (subpopulations) are required to match the observations. The two bottom panels are for the metal-poor Oosterhoff group II GCs (M15 \& M80). These GCs are similarly blue-HB dominated with a gap between the blue HB (BHB) and the extreme blue HB (EBHB), although M80 has a lower fraction of RR Lyrae stars than M15. For these GCs, a relatively large $\Delta$t(G1 - G2) is required both in synthetic HB model and chemical evolution model (see Table~\ref{tab1}).  As illustrated in detail in \citet{Jan14} and \citet{Jan15}, with such a large $\Delta$t(G1-G2), together with some $\Delta$[CNO/Fe], the color of G2 on the HB could be redder than that of G1. Therefore, in the case of M15, the instability strip (IS) is mostly occupied by He and CNO enhanced G2, reproducing a relatively long $\left<P_{\rm ab}\right>$. As shown by \citet{Jan14}, this is because the enhancement in He increases luminosity while the enhancement in CNO reduces the mass of an HB star at a given temperature, both of which increase $\left<P_{\rm ab}\right>$. Our synthetic HB model for M80 is similar to that for M15, but a smaller $\Delta$t(G1-G2) is required to match the observed HB morphology. For this GC, the IS is populated by both G1 and G2 evolved from BHB. Similarly to the result obtained from synthetic HB models, a large $\Delta$t(G1-G2) is also favored for both of these GCs from chemical evolution models. This is mostly because a large age difference allows the ejecta from low mass stars to be mixed with massive star ejecta so that oxygen is not depleted (or even somewhat enhanced) while sodium is substantially enhanced between G1 and G2. In order to reproduce EBHB, including a gap between BHB and EBHB, a sudden enhancement in He abundance between the subpopulations is required \citep[see, e.g.,][]{Dan04,Lee05}. We found that, in chemical evolution modeling, a series of starburst within a short period can reproduce such a large He enhancement with negligible variations in other light elements. For this group of subpopulations having $\Delta$t $\la$ 0.01 Gyr, $\Delta$Y $\la$ 0.001, and $\Delta$[CNO/Fe] $\la$ 0.01 between them, we denote G2-a, G2-b, and G2-c. The ejecta from these stars can then be added to form later generation stars largely enhanced in He abundance. While this series of starburst would be considered as an ad hoc assumption, this prescription is apparently required to reproduce a large He enhancement for EBHB. Direct spectroscopic observations for a difference in He abundance between BHB and EBHB stars in M15 and M80 would help to confirm this assumption.\\

\begin{figure}
\centering
\epsscale{1.1}
\plotone{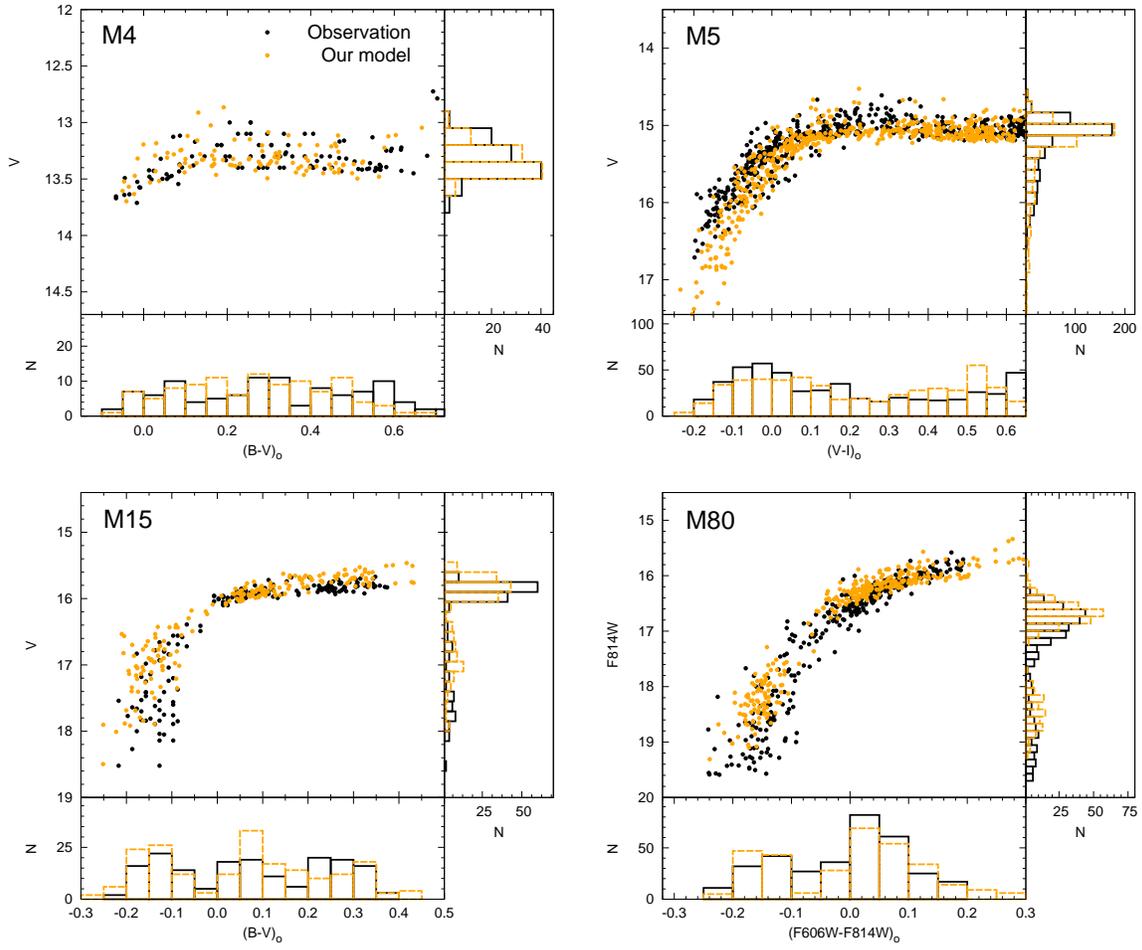}
\caption{Further comparison of our synthetic HB model with the observed HB in the color and magnitude distributions (see the text).\label{fig2}}
\end{figure}

\begin{figure}
\centering
\epsscale{0.8}
\plotone{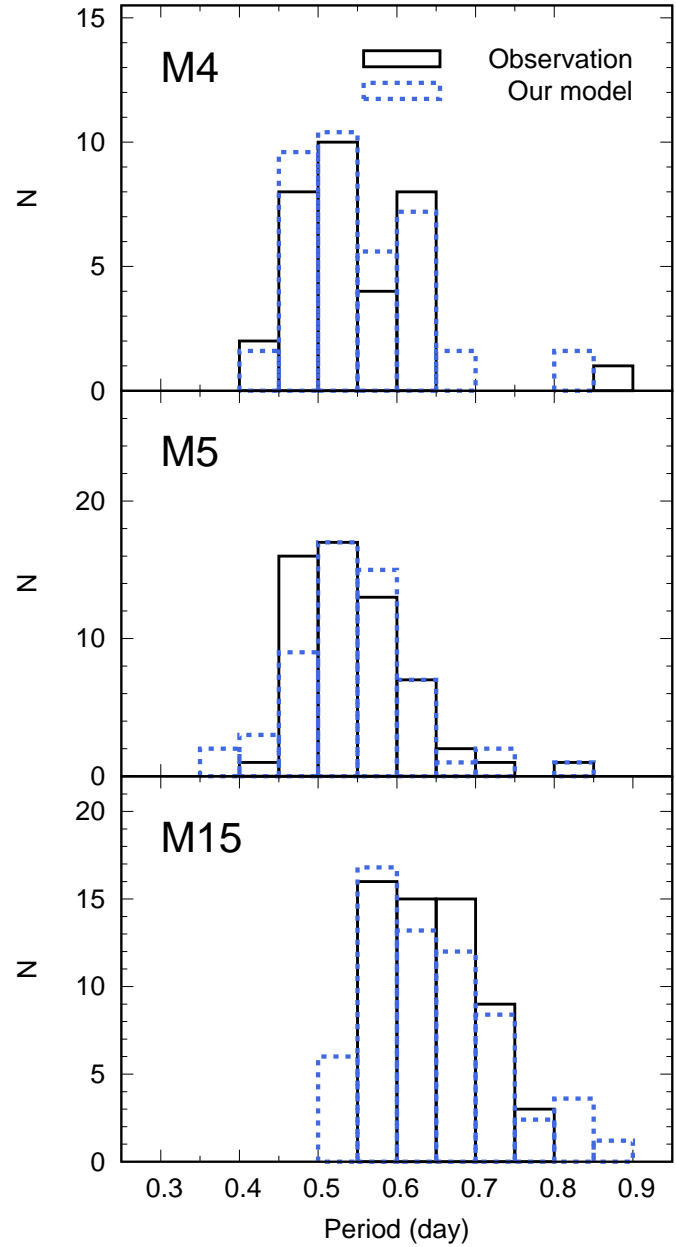}
\caption{Period distribution of type ab RR Lyrae variables from our synthetic HB model compared with that from the observation \citep[data from][]{Ste14,Are16,Cor08}. The period distribution is not shown for M80 since it has only 7 type ab RR Lyrae stars. \label{fig3}}
\end{figure}

Figure~\ref{fig2} further compares  the color and magnitude distributions between the synthetic HB model and the observation. The comparison of the color histogram shows a reasonable agreement between the HB model and photometric data. The p-values from the K-S test are also relatively large (0.12, 0.09, 0.63, and 0.064 for M4, M5, M15, and M80, respectively), indicating that the difference between the model and the observation is not statistically significant. Unlike the color distribution, the p-values for the magnitude histograms are apparently low (p $<$ 0.06), but the comparison of  the histograms still shows a reasonable agreement considering the uncertainties in HB modeling and photometry. Further fine tuning of the models would be required to obtain a better match with the observed magnitude distributions. The period distribution of type ab RR Lyrae variables from our synthetic HB model is compared with that from the observation in Figure~\ref{fig3} \citep[data from][]{Ste14,Are16,Cor08}, which shows good agreements with large p-values from the K-S test (0.99, 0.53, and 0.66 for M4, M5, and M15, respectively). For the [O/Na] histograms in the right panels of Figure~\ref{fig1}, the p-values from the K-S test are 0.87, 0.17, 0.58, and 0.64 for M4, M5, M15, and M80, respectively, confirming the agreement between our models and the observed distributions.

\section{Discussion} \label{sec:dis}

\begin{figure}
\centering
\epsscale{1.0}
\plotone{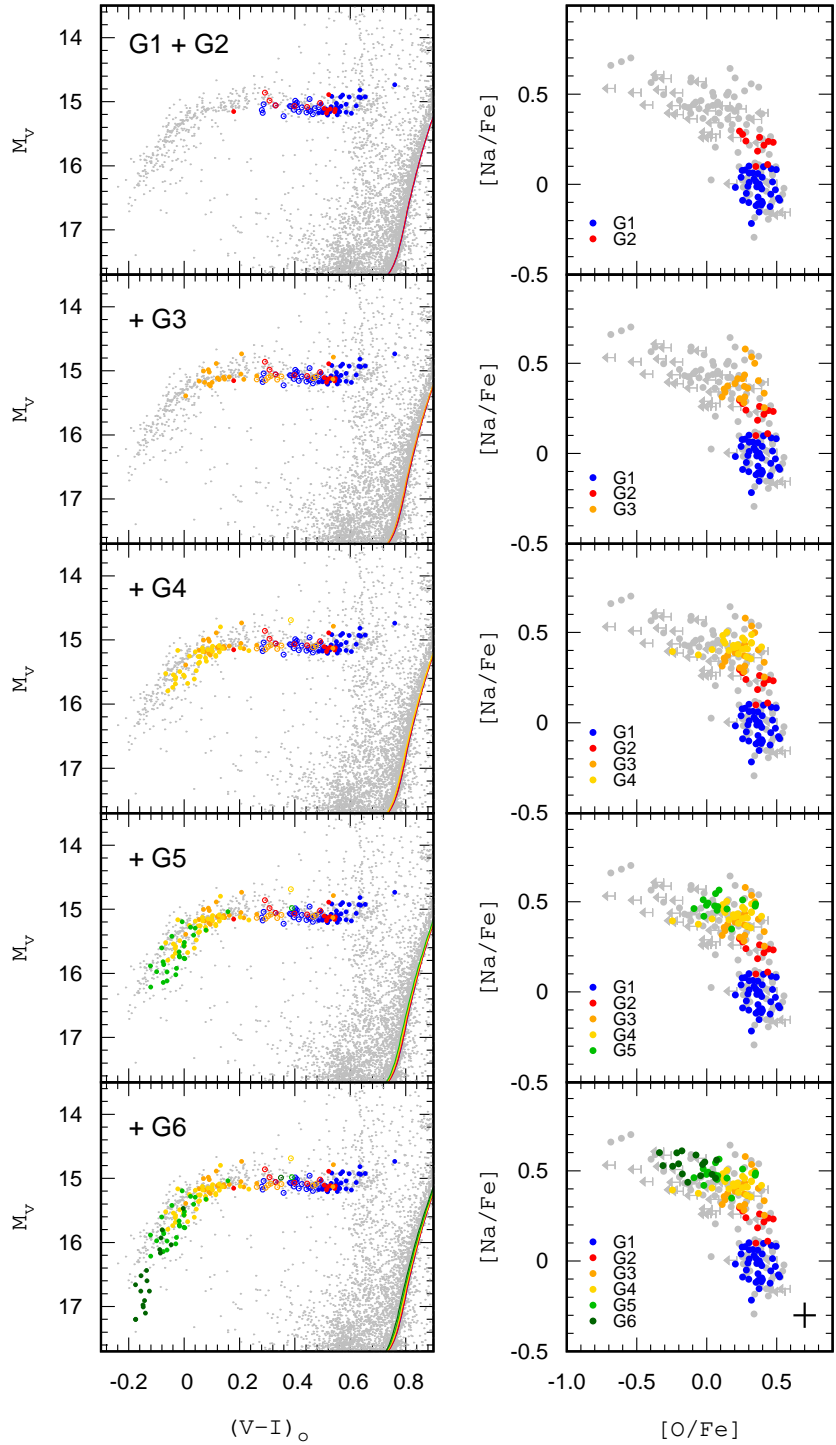}
\caption{The extensions in HB and Na-O chemical pattern in a formation sequence expected from our scenario for GCs with different number of subpopulations, based on our model for M5 as an example. The HB morphology and Na-O pattern gradually extend bluewards and upwards, respectively, as further episodes of starburst continue.\label{fig4}}
\end{figure}

In our unified models, as described above, He and Na abundances gradually increase over the generation, and therefore, the extensions in both HB morphology and Na-O chemical pattern are largely dependent on the presence of later generation stars after G2 or G3. Figure~\ref{fig4} illustrates this based on our model for M5 as an example. We can see here that the HB morphology and Na-O pattern gradually extend bluewards and upwards, respectively, as further episodes of starburst continue. Therefore, our model for M5 would be similar to that for M4 when only G1, G2, and G3 are considered both on the HB and Na-O plane. It appears from this comparison that these two GCs experienced a similar star formation history, but, in the case of M5, the formation of subpopulations lasted further to G4, G5, and G6. \\

\begin{figure}
\centering
\epsscale{1.0}
\plotone{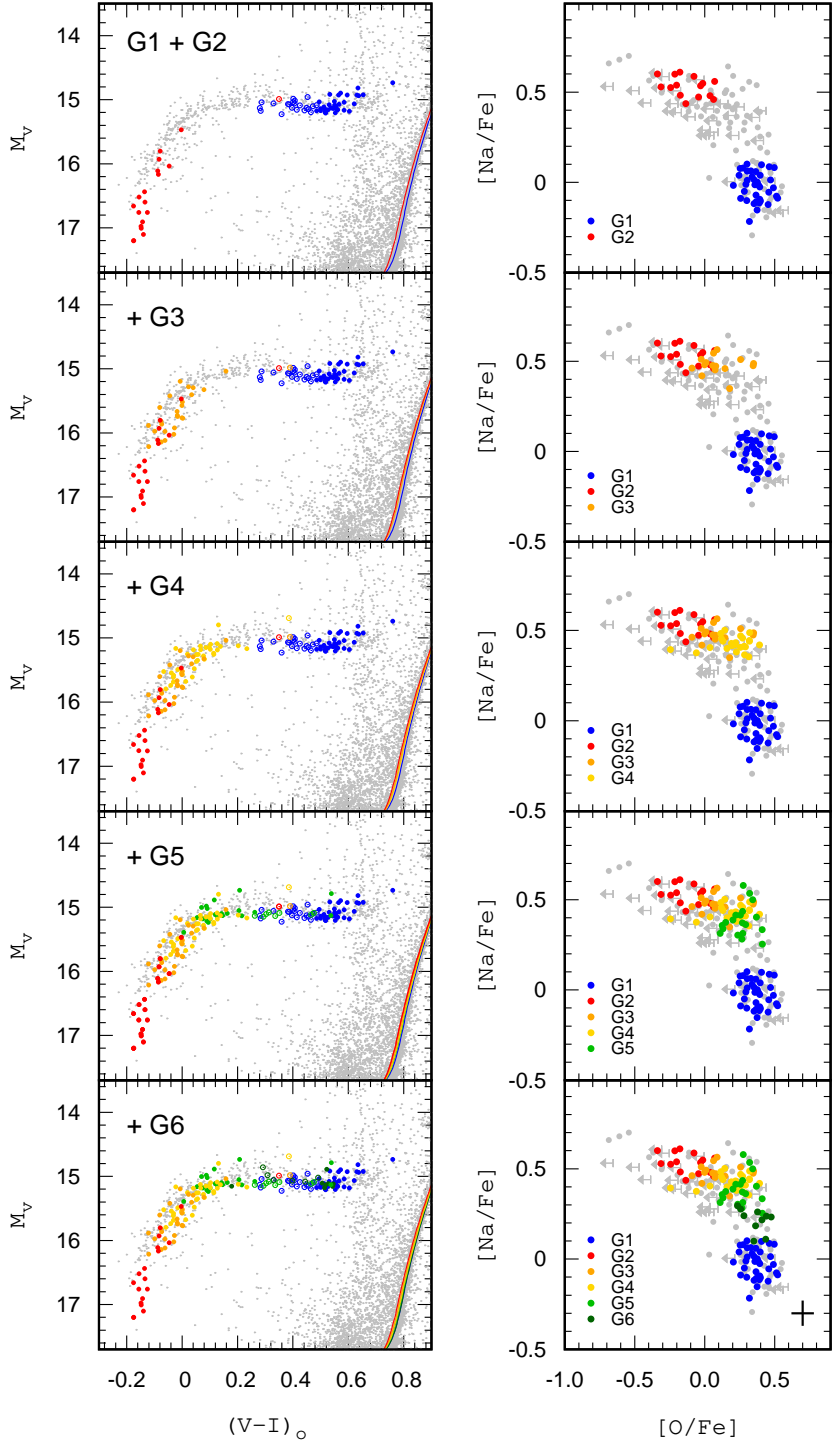}
\caption{Same as Figure~\ref{fig4}, but in a formation sequence expected from the model requiring dilution. In order to schematically show a difference with our model, the same models for M5 in Figure~\ref{fig4} are reused here for the illustration purpose. In this scenario, the G1 and G2 would be placed on the two extreme ends on both the HB and Na-O diagram, while the later generation stars, if any, would be located in between depending on the magnitude of dilution. \label{fig5}}
\end{figure}

In contrast with our model, in the scenario that requires dilution of processed material with pristine gas, the G2 forms solely from the processed material enriched in He and Na. Therefore, the G1 and G2 would be placed on the two extreme ends on both the HB and Na-O diagram, while the later generation stars, if any, would be located in between depending on the magnitude of dilution. In order to schematically show this difference between our model and the dilution scenario, for the illustration purpose, Figure~\ref{fig5} simply reuses Figure~\ref{fig4}, but in a formation sequence expected from the model requiring dilution for GCs with different number of subpopulations. It is clear from this illustration that broad gaps would inevitably appear on the HB and Na-O plane when there are only three or fewer subpopulations (see the top and second panels). Thus, in this scenario, many GCs are likely to be observed to have broad gaps on the HB and Na-O plane. However, although narrow and moderate gaps have been found in some GCs (mostly with EBHB), such a broad gap has not been observed in most cases with a possible exception of M14 on the Na-O plane \citep{Joh19}. Furthermore, as described above, both our synthetic HB models and chemical evolution models for M15 and M80 favor a relatively large $\Delta$t(G1-G2) ($\ga$ 0.5 Gyr). Such a large difference in age would be also inconsistent with some of the other models that predict a short time scale ($\la$  0.01 - 0.1 Gyr).  \\

In this paper, by simultaneously employing the stellar evolution and chemical evolution models for four GCs, we have shown that a consistent result can be obtained from these two independent studies. From this new approach, we have demonstrated that a better constraint on star formation history and chemical evolution can be placed for subpopulations in GCs. Furthermore, we have illustrated that this technic would eventually help to narrow down a probable solution among several scenarios suggested for chemical evolution in GCs. For this, more high quality spectroscopic observations are needed for a large sample of stars in GCs, those with a gap/gaps on the HB and Na-O plane in particular. In addition, theoretical and observational studies as to the SN feedback in a GC are also essential to confirm the implicit assumption made in this study.\\

We thank the anonymous referee for a number of helpful comments and suggestions. Support for this work was provided by the National Research Foundation of Korea (grants 2017R1A2B3002919 and 2017R1A5A1070354).

\clearpage

\clearpage

\clearpage

\clearpage

\clearpage

\clearpage

\begin{deluxetable*}{ccccccccc}
\tablenum{1}
\tablecaption{Best-fit parameters from synthetic HB and chemical evolution models for four GCs.\label{tab1}}
\tablewidth{0pt}

\tablehead{
\colhead{Generation} & \colhead{Y} & \colhead{$\Delta$[CNO/Fe]} & \colhead{Age (Gyr)} &\colhead{[Na/Fe]} & \colhead{[O/Fe]} & \colhead{$M_{\rm TRGB}$\tablenotemark{b}} & \colhead{$\left<M_{\rm HB}\right>$\tablenotemark{c}} & \colhead{Fraction\tablenotemark{d}}\\
\colhead{} & \colhead{$\sigma$ = $\pm$0.01} & \colhead{$\pm$0.03} & \colhead{$\pm$0.04} & \colhead{$\pm$0.08} & \colhead{$\pm$0.06} &&& \colhead{$\pm$0.03} 
}
\startdata
\multicolumn{7}{c}{M4: [Fe/H] = -1.25, $\left<P_{\rm ab}\right>$ = 0.56 (0.55), HB type\tablenotemark{a} = -0.04 (-0.07)}\\
G1 & 0.232 & 0.00 & 12.250& 0.10 & 0.30 &0.857&0.640& 0.31 \\
G2 & 0.249 & 0.11 & 12.050& 0.38 & 0.25 &0.837&0.620& 0.27 \\
G3 & 0.261 & 0.19 & 11.980& 0.52 & 0.22 &0.821&0.603& 0.42 \\
\cline{1-9}
\multicolumn{7}{c}{M5: [Fe/H] = -1.58, $\left<P_{\rm ab}\right>$ = 0.56 (0.55), HB type = 0.36 (0.39)}\\
G1 & 0.231 & 0.00 & 11.800& 0.00 & 0.35 &0.850&0.657& 0.31 \\
G2 & 0.238 & 0.07 & 11.790& 0.19 & 0.34 &0.840&0.647& 0.08 \\
G3 & 0.255 & 0.13 & 11.740& 0.35 & 0.29 &0.816&0.623& 0.16 \\
G4 & 0.264 & 0.17 & 11.735& 0.42 & 0.26 &0.803&0.610& 0.21 \\
G5 & 0.278 & 0.20 & 11.725& 0.47 & 0.18 &0.783&0.590& 0.14 \\
G6 & 0.300 & 0.31 & 11.682& 0.53 & -0.10 &0.752&0.557& 0.10 \\
\cline{1-9}
\multicolumn{7}{c}{M15: [Fe/H] = -2.16, $\left<P_{\rm ab}\right>$ = 0.64 (0.64), HB type = 0.71 (0.74)}\\
G1 & 0.230 & 0.00 & 11.780& 0.10 & 0.30 &0.841&0.679& 0.45 \\
G2-a & 0.242 & 0.54 & 10.860& 0.37 & 0.43 &0.843&0.676& 0.14 \\
G2-b & 0.243 & 0.54 & 10.855& 0.37 & 0.43 &0.841&0.675& 0.06 \\
G2-c & 0.243 & 0.55 & 10.851& 0.37 & 0.43 &0.842&0.675& 0.02 \\
G3 & 0.304 & 0.70 & 10.769& 0.47 & 0.31 &0.756&0.596& 0.18 \\
G4 & 0.306 & 0.71 & 10.766& 0.48 & 0.30 &0.753&0.593& 0.08 \\
G5 & 0.328 & 0.85 & 10.731& 0.55 & 0.24 &0.724&0.564& 0.07 \\
\cline{1-9}
\multicolumn{7}{c}{M80: [Fe/H] = -1.85, $\left<P_{\rm ab}\right>$ = 0.66 (0.68), HB type = 0.95 (0.93)} \\
G1 & 0.231 & 0.00 & 12.800& 0.15 & 0.35 &0.822&0.639& 0.21 \\
G2-a & 0.240 & 0.21 & 12.300& 0.36 & 0.37 &0.818&0.634& 0.21 \\
G2-b & 0.241 & 0.22 & 12.297& 0.36 & 0.37 &0.817&0.632& 0.09 \\
G3 & 0.249 & 0.32 & 12.296& 0.55 & 0.36 &0.805&0.622& 0.19 \\
G4 & 0.303 & 0.47 & 12.216& 0.71 & 0.22 &0.732&0.547& 0.16 \\
G5 & 0.313 & 0.52 & 12.206& 0.74 & 0.18 &0.718&0.533& 0.08 \\
G6 & 0.321 & 0.64 & 12.176& 0.79 & 0.11 &0.708&0.520& 0.06
\enddata

\tablenotetext{a}{HB type = (B-R)/(B+V+R), where B, V, and R are the numbers of blue HB stars, RR Lyrae variables, and red HB stars, respectively \citep{Lee94}. }
\tablenotetext{b}{$M_{\rm TRGB}$ is the mass (in $M_{\rm \sun}$) a star would have at the tip of the RGB if it did not lose mass.}
\tablenotetext{c}{$\left<M_{\rm HB}\right>$ is the mean mass (in $M_{\rm \sun}$) on the HB.}

\tablenotetext{d}{Population ratio is after the preferential loss of earlier generations.}
\tablecomments{For $\left<P_{\rm ab}\right>$ and HB type, observed values are in parenthesis, which are adopted from Table 1 of \citet{Jan15}.}

\end{deluxetable*}




\end{document}